# Tuning chirality amplitude at ultrafast timescales

Hiroki Ueda[1,*], Takahiro Sato[2], Quynh L. Nguyen[2], Elizabeth Skoropata[1], Ludmila Leroy[1], Tim Suter[3], Elsa Abreu[3], Matteo Savoini[3], Vincent Esposito[2], Matthias Hoffmann[2], Carl P. Romao[4,5], Julien Zaccaro[6], Diling Zhu[2], Steven Lee Johnson[1,3], and Urs Staub[1,*]

[1] *Center for Photon Science, Paul Scherrer Institute, 5232 Villigen-PSI, Switzerland.*

[2] *Linac Coherent Light Source, SLAC National Accelerator Laboratory, Menlo Park, CA, USA.*

[3]*Institute for Quantum Electronics, Physics Department, ETH Zurich, 8093 Zurich, Switzerland.*

[4]*Department of Materials, ETH Zurich, 8093 Zurich, Switzerland.*

[5]*Department of Materials, Faculty of Nuclear Sciences and Physical Engineering, Czech Technical University in Prague, Trojanova 13, Prague 120 00, Czech Republic.*

[6]*Grenoble Alpes University, CNRS, Grenoble INP, Institut Néel, BP166, 38042 Grenoble Cedex 09, France.*

[*] Correspondence authors: hiroki.ueda@psi.ch and urs.staub@psi.ch

## Abstract

Chirality is a fundamental symmetry concept describing discrete states, i.e., left-handed, right-handed, or achiral, and existing at disparate scales and in many categories of scientific fields. Even though symmetry breaking is indispensable for describing qualitatively distinct phenomena, symmetry cannot quantitatively predict measurable quantities. One can continuously distort an object, introducing the concept of "chirality amplitude", similar to representing magnetization as the "amplitude" of time-reversal symmetry breaking. Considering the role of magnetization in emergent phenomena with time-reversal symmetry breaking, chirality amplitude is intuitively a key quantity for controlling chirality-related emergent phenomena. Here, we propose two types of chiral lattice distortions and demonstrate the tunability of their amplitude in ultrafast timescales. Resonant X-ray diffraction with circular polarization is an established technique to measure crystal chirality directly. We quantify the ultrafast change in chirality amplitude in real time after an optical excitation. Using instead a THz excitation, we observe oscillations in the resonant diffraction intensities corresponding to specific phonon frequencies. This indicates the creation of additional asymmetry, which could also be described as an enhancement in chirality amplitude. Our proposed concept of chirality amplitude and its ultrafast control may lead to a unique approach to control chirality-induced emergent phenomena in ultrafast timescales.

# Main text
## Introduction

Chirality is a universal concept of defining the handedness of an object and is relevant in widely spread different scientific communities, such as biology, chemistry, and particle physics. In condensed matter physics, it has also created central topics, e.g., chiral magnetism [1], chiral fermions [2], and chiral phonons [3], which lead to emergent phenomena, such as chiral edge state [4], non-reciprocal transport [5], and phono-magnetism [6: 7: 8]. Although chirality is a basic symmetry concept underlying these phenomena, its role as an order parameter is not formulated well in comparison to, for example, magnetization in the case of time-reversal symmetry breaking. We define chirality amplitude as the amplitude of a lattice distortion associated with chiral symmetry breaking. One may argue that the amplitude of chirality in a lattice is the structural deviation from the centrosymmetric supergroup, and hence, a measure of the structural distortion contains a direct relation to the development of chirality as an order parameter or chirality amplitude. However, the direct connection between chirality amplitude and chirality-related emergent phenomena has been poorly discussed despite a substantial number of reported emergent phenomena based on this symmetry concept [9: 10: 11: 12]. The onset of chirality breaks all roto-inversion symmetries but does not necessarily induce a switchable ferroic order parameter, as is the case with magnetization. In contrast to, for example, ferromagnetic states, the absence of such a primary ferroic order switchable by a conjugate field makes it challenging to control chiral states with a few exceptions where a gyrotropic phase transition accompanies another ferroic order, e.g., ferroelectricity [13: 14: 15: 16] or ferroelasticity [17]. This is the principal reason why chirality-induced phenomena are not typically considered in terms of a chiral order parameter or chirality amplitude.

Given that chirality is an essential ingredient of various emergent phenomena, tuning the *static* crystal chirality amplitude, e.g., by chemical substitution [18], is the most straightforward and intuitive approach to control such phenomena, e.g., optical rotatory. As an asymmetric lattice distortion is often responsible for emergent spin and orbital textures [19: 20: 21], enhancing or reducing the chirality amplitude (see Fig. 1b) can affect the ground state due to unbalancing competing interactions. For example, modifying the interfacial strain on a multiferroic heterostructure can affect spin exchange coupling and result in skyrmion nucleation [22]. Yet largely unexplored to date is the *dynamical* control of crystal chirality amplitude, for which there are several possible pathways. Any phonon mode can be chiral at arbitrary momentum points in chiral crystals or polar crystals, i.e., chiral phonons [3, 23], and coherent chiral phonon excitation may affect the global crystal chirality, c.f., ferroelectric polarization reversal by driving an IR-active mode for the case of a ferroelectric order parameter [24]. Although the effect of coherent chiral phonon excitation on global chirality has not been reported yet, this is practically the same approach to induce an effective magnetic field by driving degenerate infrared- (IR) active phonon modes using a circularly polarized mid-IR or THz pulse [6: 7: 8], which can twist the lattice. Another possibility to modulate chirality at ultrafast timescales is to drive a phonon mode arising from a coexisting ferroic order that couples to a gyrotropic order, as mentioned above as exceptional cases of allowing ferroic switching of chiral order parameter [13: 14: 15: 16: 17]. Electromagnons in magnetoelectric multiferroics, magnons hybridized with polar phonons, are an example of such dynamical cross-coupling between ferroic orders [25].

In this Article, we propose two types of chiral lattice distortions to describe a chiral state and demonstrate them as tunable parameters at ultrafast timescales. Resonant X-ray diffraction with circular polarization directly measures the chirality amplitude as well as its sign, i.e., the handedness [26]. Based on this technique extended in the time domain, we investigate the ultrafast dynamics of chirality amplitude in the chiral crystal $CsCuCl_3$, triggered by an optical (400 nm) or THz pulse. In the case of the optical excitation, we find at picosecond timescales a chirality amplitude variation of ~13% with respect to the static value, demonstrating that chirality amplitude is a controllable quantity. In the case of the THz excitation, we observe oscillations in the intensity of a space-group allowed diffraction peak with defined IR-active phonon frequencies. Since the IR-active phonons create additional asymmetric lattice distortion and reduce the symmetry, this result implies that one can also enhance the chirality amplitude in ultrafast timescales by means of coherent phonon excitation.

    Chiral crystal structures belonging to the so-called Sohncke groups have space groups that contain only symmetry operations of the first kind, i.e., rotations, roto-translations, and translations. Note that not all chiral crystal structures belong to chiral space groups [27]. Among 65 Sohncke groups, 22 groups are chiral and have enantiomorphic pairs (e-p) of space groups reflecting opposite chirality (first type), e.g., $P3_121$-$P3_221$ and $P6_122$-$P6_522$. For the remaining 43 groups (second type), e.g., $P312$ and $P432$, a roto-inversion operation relates them to themselves. As a result, these groups are achiral, and there are no distinct space groups for enantiomers with opposite handedness, though the crystal structure is chiral. We call a crystal structure belonging to this type an enantiomorphically unpaired (e-u) chiral state. Asymmetry found in these types of Sohncke groups is non-handed chiral [28]. We can define an e-p chiral lattice distortion as a lattice distortion that does not change the symmetry in the first type of Sohncke groups but continuously varies chirality amplitude with defined handedness, e.g., the one shown in Fig. 1b. This type of mode belongs to the totally symmetric irreducible representation $A_1$, but it is noteworthy that not all $A_1$ modes are e-p chiral. Symmetry lowering can create another dimension of asymmetric lattice distortion, as conceptually illustrated in Fig. 1c, which, however, could be non-handed chiral. In this regard, one may argue that in addition to continuous change of an e-p chiral lattice distortion, chirality amplitude enhancement can also be achieved by further symmetry reduction, corresponding to the onset of a lattice distortion with symmetry other than $A_1$, i.e., an e-u chiral lattice distortion. A model case is a phase transition between two Sohncke groups from the first type to the second type. When a chiral lattice distortion describes a phase transition, the lattice distortion is a chiral order parameter with either e-p or e-u character.

    $CsCuCl_3$ is a hexagonal perovskite with Jahn-Teller active $Cu^{2+}$, exhibiting a phase transition from the centrosymmetric ($P6_3/mmc$) to the chiral [$P6_522$ (left-handed) or $P6_122$ (right-handed)] phases with tripling the $c$ lattice constant at $T_C = 423$ K, as seen in Fig. 1a. The transition is purely gyrotropic, meaning no onset of any trivial ferroic order, e.g., ferroelectricity or ferroelasticity [29]. Since the low-temperature phase belongs to the first type of Sohncke group, the order parameter is the e-p type lattice distortion. Multiple techniques have manifested its chiral coordination below $T_C$ by, e.g., optical rotation [30], X-ray and neutron diffraction [31], and, most directly, resonant X-ray diffraction with circular polarization [32; 33]. These experiments measured the e-p chiral order parameter defined as optical rotatory power, the X-ray or neutron diffraction intensities of superlattice reflection peaks, and the circular contrast of resonant X-ray diffraction intensities, respectively. In the last case, the circular contrast is proportional to the expectation value of a specific electric quadrupole moment [see Eqs. (1) and (2)]. Figure 1a represents the temperature dependence of the e-p chiral order parameter extracted by taking the square root of X-ray or neutron diffraction intensities of a superlattice reflection peak originating from the phase transition and optical rotatory power, shown in Ref. [31]. By heating up to $T_C$, the amplitude of the chiral order parameter decreases by ~13% compared to room temperature. By further heating, it exhibits a discontinuous jump to zero at $T_C$ because of the first-order nature of the structural phase transition. A chiral magnetic order occurs at ~10.7 K, in which asymmetric magnetic exchange interaction induced by the chiral crystal structure plays a key role [33].

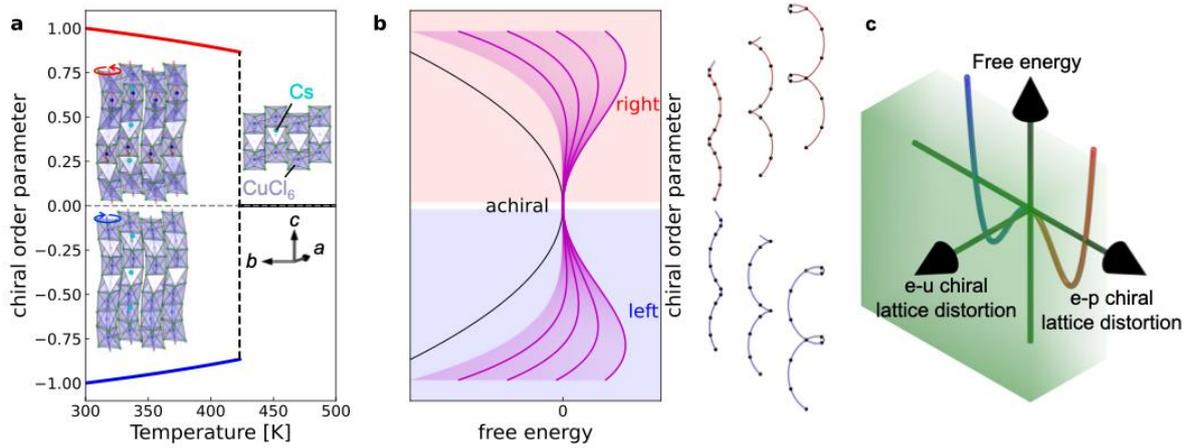

Fig. 1 | **a**, Temperature dependence of the order parameter for the chiral phase transition in CsCuCl$_3$, i.e., e-p chiral order parameter. Insets show corresponding crystal structures. Data are extracted from Ref. [31]. The order parameter is normalized to the value at 300 K. The sign of the order parameter corresponds to chirality, either left-handed (negative, blue) or right-handed (positive, red). **b**, The concept of chiral order parameter. Chirality is either there (chiral, double-well potential) or not (achiral, either single-well potential or double-well potential with thermal fluctuation). If existing, it defines the handedness, either left or right, which are e-p order parameters. However, one can continuously distort the structure to make the chirality amplitude smaller or larger (visualized by red and blue helices with different chirality amplitude on the right). **c**, Besides, another dimension representing a non-handed additional lattice distortion (e-u chiral lattice distortion) can be added when a further symmetry breaking occurs.

## Results and discussion

We performed time-resolved resonant X-ray diffraction experiments with circularly polarized X-ray pulses on an enantiopure CsCuCl$_3$ single crystal at ambient conditions (see Fig. 2 for the two experimental setups and Supplementary Materials for details). The X-ray photon energy was tuned in the vicinity of the Cu $K$ edge ($1s \to 4p$) based on the X-ray absorption spectrum (XAS), shown in Fig. 3a. At these photon energies, the resonant X-ray scattering process is sensitive to the Cu$^{2+}$ $4p$ states, which form a chiral arrangement in the chiral structure of CsCuCl$_3$.

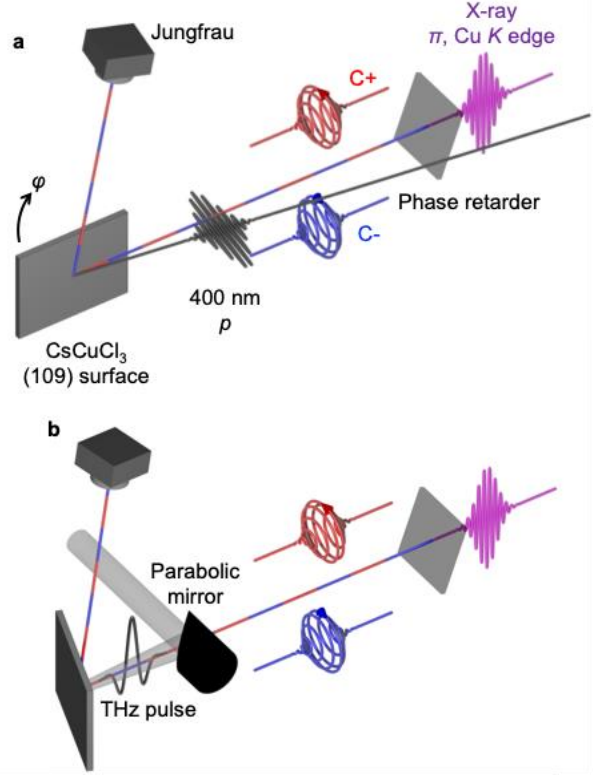

Fig. 2 | Experimental setups of pump-probe resonant X-ray diffraction with circularly polarized X-rays: **a**, Optical-pump and **b**, THz-pump. For the optical pump setup, pump (gray) and probe (red, blue, or pink, depending on polarization) beams are non-collinear with a relative angle of ~7° where X-ray beam has a grazing incidence angle of 0.5° to the sample surface in order to suppress the penetration depths mismatch. For the THz pump setup, the two beams are collinear and the incidence angle is relatively large between 40° and 70°. The electric field of the linearly polarized THz pulse was set to obtain the maximal THz electric field component along [001] on the (109) surface. A diamond-based phase retarder converts X-ray polarization from horizontal (π, purple) to circular [C+ (red) or C– (blue)]. The sample is rocked along the surface normal ($\varphi$) at a fixed incidence angle.

For the optical pump experiments, we focused on the (0 0 10) space-group forbidden reflection due to its direct sensitivity to crystal chirality, as manifested by the circular contrast in its intensity shown in Fig. 3**b** and also in Ref. [32]. As described in Supplementary Materials and Ref. [33], the (0 0 10) forbidden reflection arises purely from electric quadrupole (orbital) scattering and has circular-polarization dependent diffraction intensities described as

$$I_{(0\,0\,10)}^{P_2} = \frac{27}{8}|\langle Q_{x^2-y^2}\rangle|^2(1+\sin^2\theta)(1+\chi P_2\sin\theta)^2 \text{ and} \quad (1)$$

$$I_{(0\,0\,10)}^{+1} - I_{(0\,0\,10)}^{-1} \propto \chi|\langle Q_{x^2-y^2}\rangle|^2. \quad (2)$$

Here, $\theta$ is the Bragg angle, $P_2$ (= ±1 for C+ and C– circular polarizations) denotes the circular polarization state of an incident X-ray beam, $\chi$ represents crystal chirality (±1 for right- and left-handedness, see Fig. 1**a**), and $Q_{x^2-y^2}$ is a local $Cu^{2+}$ $4p$ electric quadrupole moment with $x$ being along the twofold axis <110> of the site and $y$ being normal to both $x$ and [001]. The circular contrast in the diffraction intensities is proportional to chirality $\chi$ and the square of the expectation value of the electric quadrupole moment $\langle Q_{x^2-y^2}\rangle$, which becomes zero in the centrosymmetric phase. Namely, the amplitude of circular contrast measures the amplitude of the e-p chiral order parameter $|\langle Q_{x^2-y^2}\rangle|$, and the sign of the circular contrast defines the handedness of the chiral crystal structure. Unlike trigonal crystals [26], the diffraction intensity formulated as Eq. (1) is independent of the azimuthal angle along [001]. This is because the azimuthal angle dependence in trigonal crystals appears through the interference of scatterings by two electric quadrupole moments while there is only a single electric quadrupole moment contributing to the (0 0 10) diffraction intensity in the hexagonal $CsCuCl_3$. Note that this resonantly allowed space-group forbidden reflection in the chiral phase becomes forbidden, even at resonance, once the crystal transforms to the centrosymmetric phase.

For the THz pump experiments, we additionally measured the (1 0 20) space-group allowed Bragg reflection, which also shows a circular contrast at resonance, as shown in Fig. 3c. The (1 0 20) reflection also appears as a non-resonant superlattice reflection only below the chiral phase transition. Therefore, this reflection intensity measures the structural distortion from the centrosymmetric phase (charge scattering) in addition to the size of the $Cu^{2+}$ 4$p$ electric quadrupole moment, which resonantly contributes to the diffraction intensities and may give rise to the interference effect with the charge scattering [34]. Modulation of this diffraction intensity in the absence of any variation in the (0 0 10) forbidden reflection intensity indicates a measure of an e-u chiral lattice distortion. Note that, in general, the circular contrast requires finite intensities in the polarization rotation channel, i.e., σ–π' and/or π–σ'. This means that the orbital scattering contributions must be involved to obtain a circular contrast because charge scattering is isotropic and cannot create X-ray polarization rotations by itself. However, the existence of an interference term between the orbital and charge scatterings can transfer changes in the charge scattering amplitude, including the onset of an e-u chiral lattice distortion, into changes of circular contrast even in the absence of an electric quadrupole modulation (e-p chiral order parameter).

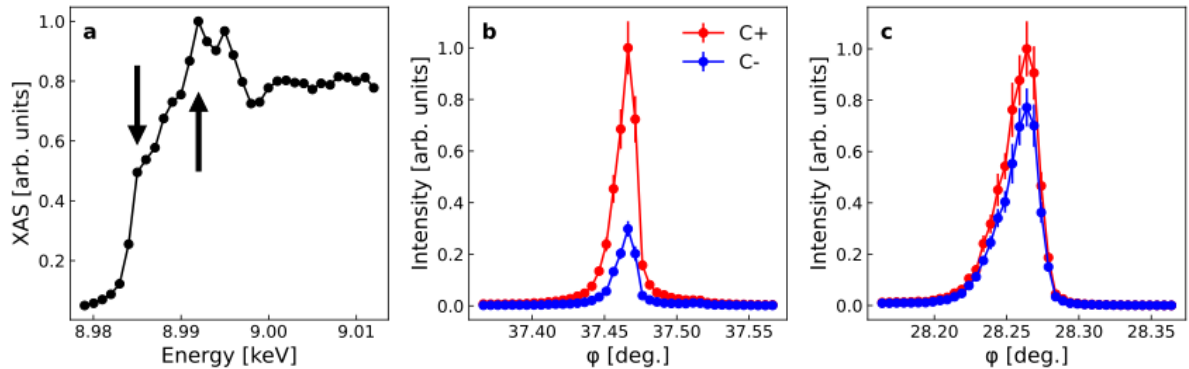

Fig. 3 | **a**, XAS around the Cu $K$ edge. The black arrows represent the photon energies employed for the diffraction measurements [8.985 keV for the (0 0 10) reflection and 8.992 keV for the (1 0 20) reflection]. **b**,**c**, Rocking curves with circular polarization around **b** the (0 0 10) forbidden reflection and **c** the (1 0 20) allowed reflection. Error bars are standard deviation.

**Optical excitation**

Figure 4a shows time traces of the (0 0 10) reflection intensity with both circular polarizations following the above bandgap optical excitation (~30 mJ/cm²) [35]. In addition to the clear intensity reduction in each polarization channel, the circular contrast also decreases. The circular contrast is directly proportional to the square of the expectation value of the electric quadrupole moment (see Eq. 2), which represents here the e-p chiral order parameter. This reduction implies that the chirality amplitude has decreased by ~13% in the probed region at 38 ps after the optical excitation. The comparison to the equilibrium case of the e-p chiral order parameter shown in Fig. 1a allows us to assign an average temperature of the probed region that almost reaches $T_C$ at 38 ps after the photoexcitation. Since there is a relatively large mismatch in the penetration depths between the pump (~26 nm) and probe (~138 nm) beams [36, 37] even at the grazing incidence, most of the energy is deposited at the surface and dissipates into deeper-lying regions on slower timescales. This behavior is consistent with the observed slow dynamics of the e-p chiral order parameter shown in Fig. 4b. The deposited energy density at the surface is much higher than that needed to bring the sample above $T_C$, which transforms the surface to the centrosymmetric phase, as found in Fig. 1a. In the recovery process to the low-temperature chiral phase, the crystal chooses between one of the handedness (left- or right-handed). The absence of any clear change in diffraction intensity for both circular polarization channels over the repetition of the data acquisition indicates that the recovery process deterministically chooses the original chiral state. This is in contrast to the global phase transition happening via a thermal cycle across $T_C$, which stabilizes a racemic state [31]. This is most likely due to the fact that the remaining deeper-lying region of the crystal acts like a seed, in contrast to the case of the thermal cycle, where chiral nucleation with random handedness leads to a racemic state.

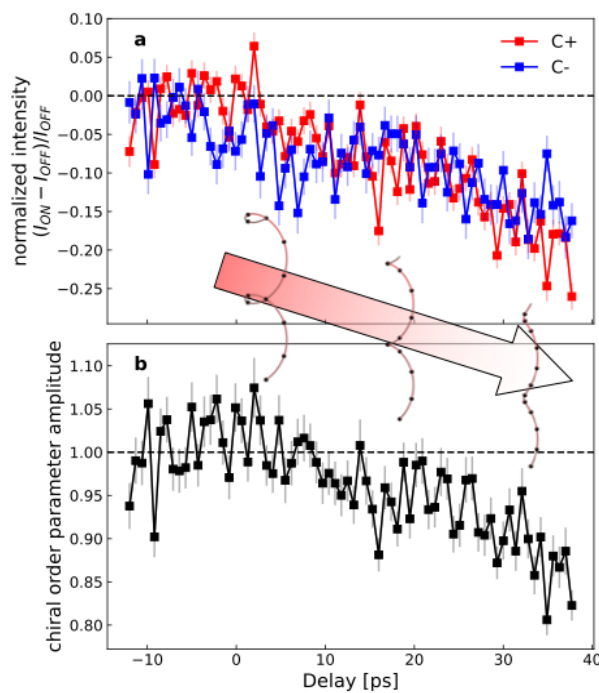

Fig. 4 | Time trace of the chiral order parameter following the optical excitation. **a**, Normalized intensity $(I_{ON} - I_{OFF})/I_{OFF}$ for the (0 0 10) forbidden reflection with two polarization C+ (red) and C– (blue), and **b**, the e-p chiral order parameter obtained by taking the square root of the circular contrast signal $(I_{C+} - I_{C-})_{ON}/(I_{C+} - I_{C-})_{OFF}$. The origin of the horizontal axis is when pump and probe beams temporary overlap on the sample. Error bars are standard deviation.

**THz excitation**

The blue plot in Fig. 5a represents the transmitted THz intensity spectrum through a thin $CsCuCl_3$ crystal (~500 μm) measured by THz time-domain spectroscopy (see Supplementary Materials for details). The employed geometry is sensitive to IR-active phonon modes with in-plane polarity, i.e., doubly degenerate $E_1$ modes. The strong absorption around 2–2.5 THz is consistent with the previously reported $E_1$ modes [38], whose frequencies are highlighted by light-red bars and tabulated in Table S1 in Supplementary Materials. The presence of the $E_1$ modes in this frequency range is supported by our DFT calculations (see Table S2 in Supplementary Materials). The black curve in Fig. 5a shows the fast Fourier transform spectrum of the THz electric field measured by electro-optic sampling (EOS) and used to drive IR-active modes in the time-resolved resonant X-ray diffraction experiment. The spectral range matches well some doubly degenerate $E_1$ (light-red bars) and non-degenerate IR-active $A_2$ (light-blue bars) phonon modes.

A mode directly modifying the e-p chiral order parameter must belong to the totally symmetric irreducible representation $A_1$. The (0 0 10) reflection is directly sensitive to the e-p order parameter modifications, but there is no detectable variation of its intensity following a THz pulse (see Fig. S2). Therefore, the e-p chiral order parameter does not strongly couple to the THz pulse resonant to the IR-active phonons.

Figures 5c and 5d show the time traces of the (1 0 20) Bragg reflection intensity with opposite circular polarizations and the corresponding circular contrast, respectively. The THz pulse evidently modulates the diffraction intensities. Their fast Fourier transform spectra plotted in Fig. 5b show multiple peaks. The prominent peaks are close to the IR-active $A_2$ mode(s) at ~1.5 THz and the IR-active $E_1$ modes at ~2 THz and ~2.5 THz [38]. These polar eigenmodes break the equilibrium non-polar symmetry of $P6_{1(5)}22$, meaning the onset of e-u chiral lattice distortions (see Fig. 5e for the concept). Note that since all phonon modes in the chiral symmetry except for $A_1$ modes lower the symmetry of the parent chiral phase, their onset corresponds to the development of an e-u chiral lattice distortion regardless of whether they are IR-active or not.

The peak amplitude in the fast Fourier transform spectra might exhibit certain dichroism (see, for example, the peaks at ~1.5 THz and ~2 THz in Fig. 5b). This can be explained when the orbital scattering terms and/or the pure charge scattering that interferes with the orbital scattering show modulations. Both cases could result in the observed circular contrast in the fast Fourier transform spectra. Note that in equilibrium, the (1 0 20) reflection has an additional sensitivity to $\langle Q_{3z^2-r^2} \rangle$ than the (0 0 10) reflection, which is solely sensitive to $\langle Q_{x^2-y^2} \rangle$. Here, the local $z$ axis is parallel to [001].

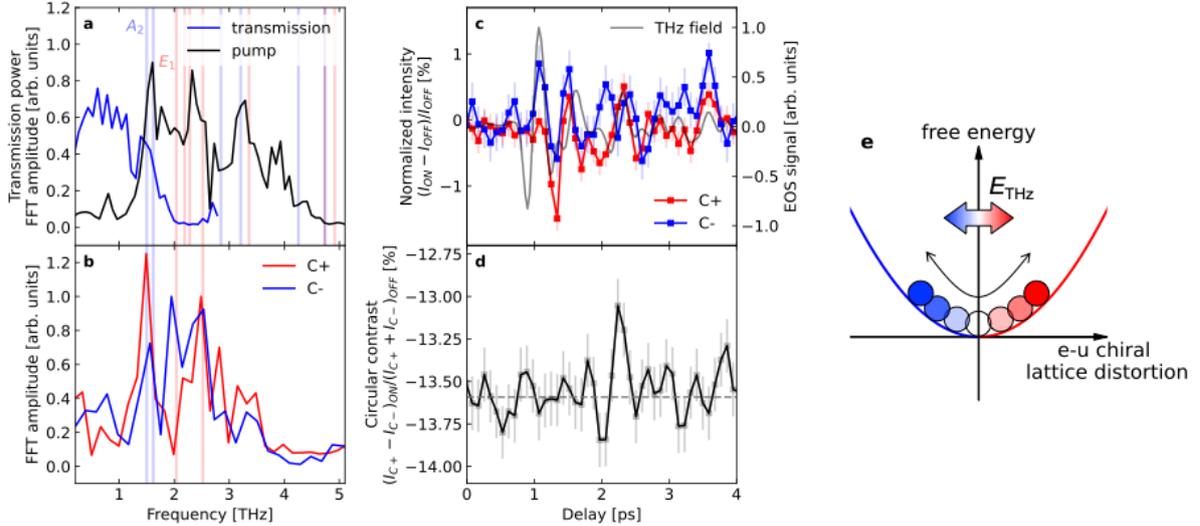

Fig. 5 | THz pulse-triggered dynamics. **a**, THz spectra of transmission power (blue) and electric field used as a pump beam in the time-resolved resonant X-ray diffraction experiment (black). Light-color bars represent IR-active $A_2$ (blue) or $E_1$ (red) phonon modes reported in Ref. [38]. **b**, Fast Fourier transform spectra of the time traces of the (1 0 20) reflection intensities [red (C+) and blue (C−)]. Selected relevant $A_2$ (blue) and $E_1$ (red) phonon frequencies are represented by light-color bars as **a**. (c,d) Time traces of the diffraction intensities of the (1 0 20) allowed reflection with circular polarization: **c**, Normalized intensities ($I_{ON} - I_{OFF}$)/$I_{OFF}$ for C+ and C−, and **d**, corresponding circular contrast. Gray curve in **c** represents the time trace of a THz electric field of the pump. Error

bars are standard deviation. **e**, A schematic of the emerged e-u chiral lattice distortion oscillating around the equilibrium by a THz electric field.

When an IR-active phonon mode is excited in a non-polar crystal, like CsCuCl$_3$ with the space group $P6_{1(5)}22$, a temporal snapshot (within a cycle of the frequency) of the crystal structure possesses a macroscopically developed electric dipole moment that does not exist in equilibrium, resulting in a lower (polar) symmetry. In the time average, the symmetry remains the same. The maximum polar subgroups of $P6_{1(5)}22$ are $P6_{1(5)}$ for an out-of-plane moment ($A_2$) and $C2$ for an in-plane moment ($E_1$). The oscillatory polarity lowers the symmetry and corresponds to the onset of an e-u chiral lattice distortion by adding another dimension in the chiral order parameter space (see Fig. 1**c** and Fig. 5**e**).

Our experimental data support the possible presence of two distinct types of chiral lattice distortions in crystal structures in the excited state as described by the second type of Sohncke groups, i.e., e-p and e-u, and the controllability of these two chiral lattice distortions in ultrafast timescales depending on the type of the drive. The argument of two distinct types of chiral lattice distortions can be generalized even in equilibrium. Since the e-p chiral order parameter obtained from the (0 0 10) reflection, as for data displayed in Fig. 4, does not show any clear temporal modulation (see Fig. S2), the observed oscillations on the (1 0 20) reflection are, therefore, likely due to the onset and oscillation of an e-u chiral lattice distortion in both amplitude and sign (see Fig. 5**e**). This demonstrates a tuning of the e-u chirality amplitude in ultrafast timescales by pushing the state along a parameter axis that is absent in equilibrium, in contrast to the optical excitation case, which alters the e-p chirality amplitude.

Very recently, the induction of macroscopic chirality in a non-chiral crystal possessing compensated chiral sublattices with opposite handedness, namely, an antiferro-chiral state, has been reported based on nonlinear phononic coupling, i.e., driving an IR-active mode coupling to a Raman-active mode that lifts off the degeneracy of two chiral sublattices and results in a second-type Sohncke group [39, 40]. This corresponds to a phononic displacement of a state along the e-u chiral parameter axis. Via the same technique, one may be able to coherently displace an e-p chiral order parameter unidirectionally, which corresponds to diagonal displacements in the e-p and e-u chiral parameter space (see Fig. 1**c**).

Tuning chirality amplitude by light and decomposing it into e-p and e-u chiral lattice distortions, as demonstrated here, could extract its hidden but critical role in chirality-related emergent phenomena. The concept of two types of chiral lattice distortions is not limited to dynamic cases but could be extended to static cases. It may be especially useful when dealing with a chiral crystal belonging to a second type Sohncke group because the role of an e-p and e-u chiral lattice distortions could be qualitatively different. Identifying the key parameter helps in understanding emergent phenomena and designing materials. One could identify the dimension in the chiral parameter space and push the state into a specific direction. Directly tuning interactions associated with a chiral lattice distortion by coherently exciting specific phonon modes, e.g., electromagnons [41, 42] and chiral soft phonons [43], may also provide a platform to investigate the role of chirality amplitude in physics. An intense THz pulse has the potential for macroscopic emergence or coherent excitation of the chirality degree in a phononic system. In relation to being developed "chiral" phononics [3; 6; 7; 8; 23], the approaches shown in our study to tailor chirality at ultrafast timescales may lead to thermally non-accessible states and related emergent phenomena in chiral crystals or compensated chiral crystals.

## Acknowledgments

We acknowledge Prof. Motohiro Suzuki from Kwansei Gakuin University for his help in developing the diamond-based phase retarder. We thank B. Pedrini for his support of static characterization. The time-resolved X-ray diffraction experiments were carried out at the XPP beamline in the Linac Coherent Light Source (LCLS), SLAC National Accelerator Laboratory. Use of LCLS is supported by the U.S. Department of Energy, Office of Science, Office of Basic Energy Sciences under Contact No. DE-AC02-76SF00515. H.U. was partially supported by the National Centers of Competence in Research in Molecular Ultrafast Science and Technology (NCCR MUST-No. 51NF40-183615) from the Swiss National Science Foundation and from the European Union's Horizon 2020 research and innovation program under the Marie Skłodowska-Curie Grant Agreement No. 801459 – FP-RESOMUS. Q.L.N. acknowledges support from the Bloch Fellowship in Quantum Science and Engineering by the Stanford-SLAC Quantum Fundamentals, Architectures, and Machines Initiative. E.S. received funding from the European Union's Horizon 2020 research and innovation programme under the Marie Skłodowska-Curie grant agreement No 884104 (PSI-FELLOW-III-3i). C.P.R acknowledges support from the project FerrMion of the Ministry of Education, Youth and Sports, Czech Republic, co-funded by the European Union (CZ.02.01.01/00/22_008/0004591), the European Union and Horizon 2020 through grant no. 10103035, and ETH Zurich. Computational resources were provided by the Swiss National Supercomputing Center (CSCS) under project ID eth3.

## Author contributions

H.U. and U.S. proposed this scientific project. J.Z. grew a single crystal. H.U., T.Su., and E.A. performed THz time-domain spectroscopy. T.Sa. implemented the circular polarization X-ray setup. H.U., T.Sa., Q.L.N., M.H., D.Z., and U.S. performed the optical pump part of the time-resolved X-ray diffraction experiments, and H.U., T.Sa., Q.L.N., M.H., D.Z., E.S., L.L., T.Su., E.A., M.S., S.L.J., and U.S. performed the THz pump part of the time-resolved X-ray diffraction experiments. H.U. and Q.L.N. analyzed the experimental data with inputs from T. Sa. C.P.R. performed DFT calculations. H.U. and U.S. wrote the manuscript with contributions from all authors.

**Competing interests**: The authors declare no competing interests.

## Data availability

Experimental and model data are accessible from the PSI Public Data Repository [44].